\PassOptionsToPackage{unicode}{hyperref}
\PassOptionsToPackage{hyphens}{url}
\PassOptionsToPackage{dvipsnames,svgnames,x11names}{xcolor}
\documentclass[
]{article}
\usepackage{amsmath,amssymb}
\usepackage{lmodern}
\usepackage{iftex}
\ifPDFTeX
  \usepackage[T1]{fontenc}
  \usepackage[utf8]{inputenc}
  \usepackage{textcomp} 
\else 
  \usepackage{unicode-math}
  \defaultfontfeatures{Scale=MatchLowercase}
  \defaultfontfeatures[\rmfamily]{Ligatures=TeX,Scale=1}
\fi
\IfFileExists{upquote.sty}{\usepackage{upquote}}{}
\IfFileExists{microtype.sty}{
  \usepackage[]{microtype}
  \UseMicrotypeSet[protrusion]{basicmath} 
}{}
\makeatletter
\@ifundefined{KOMAClassName}{
  \IfFileExists{parskip.sty}{%
    \usepackage{parskip}
  }{
    \setlength{\parindent}{0pt}
    \setlength{\parskip}{6pt plus 2pt minus 1pt}}
}{
  \KOMAoptions{parskip=half}}
\makeatother
\usepackage{xcolor}
\usepackage{graphicx}
\makeatletter
\def\maxwidth{\ifdim\Gin@nat@width>\linewidth\linewidth\else\Gin@nat@width\fi}
\def\maxheight{\ifdim\Gin@nat@height>\textheight\textheight\else\Gin@nat@height\fi}
\makeatother
\setkeys{Gin}{width=\maxwidth,height=\maxheight,keepaspectratio}
\makeatletter
\def\fps@figure{htbp}
\makeatother
\providecommand{\tightlist}{%
  \setlength{\itemsep}{0pt}\setlength{\parskip}{0pt}}
\NewDocumentCommand\citeproctext{}{}
\NewDocumentCommand\citeproc{mm}{%
  \begingroup\def\citeproctext{#2}\cite{#1}\endgroup}
\makeatletter
 \let\@cite@ofmt\@firstofone
 \def\@biblabel#1{}
 \def\@cite#1#2{{#1\if@tempswa , #2\fi}}
\makeatother
\newlength{\cslhangindent}
\newlength{\csllabelwidth}
\newenvironment{CSLReferences}[2] 
 {\begin{list}{}{%
  \setlength{\itemindent}{0pt}
  \setlength{\leftmargin}{0pt}
  \setlength{\parsep}{0pt}
  \ifodd #1
   \setlength{\leftmargin}{\cslhangindent}
   \setlength{\itemindent}{-1\cslhangindent}
  \fi
  \setlength{\itemsep}{#2\baselineskip}}}
 {\end{list}}
\usepackage{calc}

\newcommand{\pandocbounded}[1]{#1}
\ifLuaTeX
\usepackage[bidi=basic]{babel}
\else
\usepackage[bidi=default]{babel}
\fi
\babelprovide[main,import]{american}

\def\languageshorthands#1{}
\ifLuaTeX
  \usepackage{selnolig}  
\fi
\IfFileExists{bookmark.sty}{\usepackage{bookmark}}{\usepackage{hyperref}}
\IfFileExists{xurl.sty}{\usepackage{xurl}}{} 
\hypersetup{
  pdftitle={Newtrinos.jl: A Julia Package for Global Analysis of
Neutrino Data},
  pdfauthor={Philipp Eller, David Schultheiß},
  pdflang={en-US},
  colorlinks=true,
  linkcolor={Maroon},
  filecolor={Maroon},
  citecolor={Blue},
  urlcolor={Blue},
  pdfcreator={LaTeX via pandoc}}

\title{Newtrinos.jl: A Julia Package for Global Analysis of Neutrino
Data}

\definecolor{c53baa1}{RGB}{83,186,161}
\definecolor{c202826}{RGB}{32,40,38}

\usepackage[affil-it]{authblk}
\usepackage{orcidlink}
\author[1%
  ]{Philipp Eller%
    \,\orcidlink{0000-0001-6354-5209}\,%
    }
\author[1%
  ]{David Schultheiß%
    \,\orcidlink{0009-0000-3027-684X}\,%
    }

\affil[1]{Technical University of Munich, Germany%
  }
\date{28 October 2025}

\begin{document}
\maketitle

\section{Summary}\label{summary}

\emph{Newtrinos.jl} is an open-source Julia package for performing
global analyses of neutrino data. It provides a modular, three-layer
architecture that separates physics models, experiment descriptions, and
statistical inference into independent modules. This allows researchers
to freely combine experiments and test them against a variety of
theoretical models. New experiments and physics models can be added
without modifying core code.

Full statistical forward models including all relevant systematic
uncertainties are implemented for each experimental dataset, defining
both the likelihood and the data-generating process and enabling modern
statistical inference workflows.

The framework is composable, making it straightforward to construct
joint likelihoods over multiple datasets. Physics and nuisance
parameters can be merged, correlated, or decorrelated across experiments
to ensure consistency in the joint fit.

The package supports Frequentist and Bayesian inference and is
parallelizable across CPU threads or distributed workers. Written
entirely in Julia, all models are automatically differentiable, enabling
exact gradient computation.

\section{Statement of Need}\label{statement-of-need}

As neutrino physics transitions into an era of high-precision
measurements, global fits have become essential for extracting neutrino
properties and probing physics beyond the Standard Model
(\citeproc{ref-Capozzi:2025wyn}{Capozzi et al., 2025};
\citeproc{ref-Esteban:2024eli}{Esteban et al., 2024};
\citeproc{ref-deSalas:2020pgw}{Salas et al., 2021}). By combining
disparate datasets, they break parameter degeneracies that limit
single-experiment analyses. Consequently, they are poised to address
urgent open questions such as the neutrino mass ordering and CP
violation in the lepton sector. The first joint fit between T2K and NOvA
demonstrated that combining distinct experimental signatures is crucial
to resolving such degeneracies and tensions
(\citeproc{ref-T2K:2025wet}{Abubakar \& others, 2025}).

Performing a global neutrino fit requires assembling diverse,
experiment-specific detector models, neutrino flux models, matter
density profiles, interaction cross sections, et cetera, into a coherent
inference pipeline, while rigorously accounting for large numbers of
correlated systematic uncertainties. Varying availability of public data
and documentation further limits dataset compatibility, making global
fits a substantial undertaking that can render certain analyses
infeasible (\citeproc{ref-Capozzi:2025wyn}{Capozzi et al., 2025}).
Existing frameworks that have enabled successful studies are typically
proprietary and closed-source, making results difficult to verify and
the software impossible to reuse or adapt.

A further challenge is the computational burden of high-dimensional
parameter spaces with correlated systematics. As the number of
experiments grows, traditional statistical methods become increasingly
intractable. Automatic differentiation and gradient-based inference
offer promising paths forward. However, existing frameworks (see Related
Work) were developed before these techniques became widely adopted and
therefore rely primarily on derivative-free optimization or
finite-difference gradient estimates such as those used by MINUIT
(\citeproc{ref-James:1975dr}{James \& Roos, 1975}).

\emph{Newtrinos.jl} addresses these limitations with a fully
open-source, extensible framework built within the high-performance
Julia ecosystem.

\section{Key Features}\label{key-features}

Key features include:

\begin{itemize}
\item
  \textbf{Out-of-the-box usability}: The package includes experimental
  data, configuration files, and plotting tools, making it
  straightforward to reproduce results, verify correctness, and conduct
  custom analyses on existing models and experiments.
\item
  \textbf{Modular architecture}: Experiments, physics models, and
  analysis methods are implemented as independent components with
  clearly separated concerns. The physics layer provides theory
  predictions and physical parameters; the experiment layer configures
  individual experiments with their parameters, priors, and forward
  models; the analysis layer provides inference methods for sampling,
  profiling, or scanning. Each layer treats the others as black boxes,
  enabling flexible composition. Layers communicate through two
  well-defined interfaces: NamedTuples of parameters and priors adhering
  to the Distributions.jl standard, and callable functions stored in
  structs. Theory extensions can be added via Julia's multiple dispatch
  without modifying existing code, and parameters flow through the
  pipeline automatically.
\item
  \textbf{Full statistical models}: The package constructs full
  statistical forward models for each experiment, encoding the complete
  mapping from physical parameters to expected observations. Systematic
  uncertainties enter as nuisance parameters rather than being absorbed
  into simplified \(\chi^2\) approximations, preserving the full
  statistical information. This enables profiling or marginalisation of
  joint likelihoods, Bayesian posterior estimation, and the generation
  of pseudo-data and prior and posterior predictive checks.
\item
  \textbf{Consistent parameter handling}: All physics and nuisance
  parameters across configured experiments are collected into a single,
  flat NamedTuple via \texttt{get\_params} and \texttt{get\_priors} with
  \texttt{safe\_merge}. Wrapper functions support arbitrary parameter
  renaming and (de)correlation between experiments. Prior distributions,
  including correlated priors via covariance matrices, are managed in
  the same unified structure and compose naturally into a joint prior.
  Individual parameters can be overridden or conditioned at analysis
  time without modifying the underlying modules, keeping configurations
  transparent and reproducible.
\item
  \textbf{Automatic differentiation}: The complete forward model chain
  --- from oscillation probability calculation to likelihood evaluation
  --- is compatible with \emph{ForwardDiff.jl} out of the box, with
  Mooncake.jl also successfully tested. Exact, efficient gradients
  enable advanced optimisation and inference methods such as those
  provided by (\citeproc{ref-Schulz:2021BAT}{Schulz et al., 2021}).
\item
  \textbf{Scalability}: The package is built for large-scale inference
  tasks with tens of experiments and hundreds of parameters. Julia's
  just-in-time compilation delivers high performance without sacrificing
  expressiveness. Profile likelihood scans parallelise transparently
  across CPU threads or distributed workers via Distributed.jl, and
  forward-mode automatic differentiation keeps gradient-based inference
  tractable as the parameter space grows.
\end{itemize}

\section{Workflow Overview}\label{workflow-overview}

\emph{Newtrinos.jl} is designed as a modular pipeline where components
communicate strictly through functional interfaces: physics modules
expose callables used inside experiment forward models, and experiments
expose forward model callables consumed by the analysis layer, without
either inspecting the other's internals. A typical analysis proceeds as
follows:

\begin{enumerate}
\def\labelenumi{\arabic{enumi}.}
\item
  \textbf{Configure physics}: Select and instantiate theoretical models
  --- for example, a three-flavour oscillation model with non-standard
  matter interactions, an Earth density model, an atmospheric flux
  model, and a cross-section model. Any model can be swapped
  independently, e.g.~replacing a three-flavour model with a BSM model
  including sterile neutrinos.
\item
  \textbf{Configure experiments}: Select one or more experiment modules,
  each encapsulating its detector response, systematic uncertainties,
  and observed data. Instantiate them via the \texttt{configure} method
  with the chosen physics model. Experiments are independent and can be
  freely combined, typically sharing a common physics model.
\item
  \textbf{Build the joint likelihood}: Pass the collection of configured
  experiments to \texttt{generate\_likelihood}, which composes their
  individual forward models and likelihood functions into a single joint
  likelihood.
\item
  \textbf{Collect parameters and priors}: Use \texttt{get\_params} and
  \texttt{get\_priors} to collect and merge all physics and nuisance
  parameters across experiments into a unified NamedTuple. Parameters
  and priors can be modified as needed, e.g.~for likelihood conditioning
  via Accessors.jl.
\item
  \textbf{Run inference}: Choose an analysis method from the provided
  tools. For example, run a profile likelihood scan by calling
  \texttt{profile} on the joint likelihood over a chosen parameter grid
  to obtain a \texttt{NewtrinosResult} containing the grid coordinates,
  likelihood values, optimised nuisance parameters at each point, and
  run metadata.
\item
  \textbf{Visualize and export results}: Plot confidence contours or
  best-fit data/MC comparisons using the built-in plotting utilities
  based on Makie.jl, and save results to disk.
\end{enumerate}

\begin{figure}
\centering
\pandocbounded{\includegraphics[keepaspectratio]{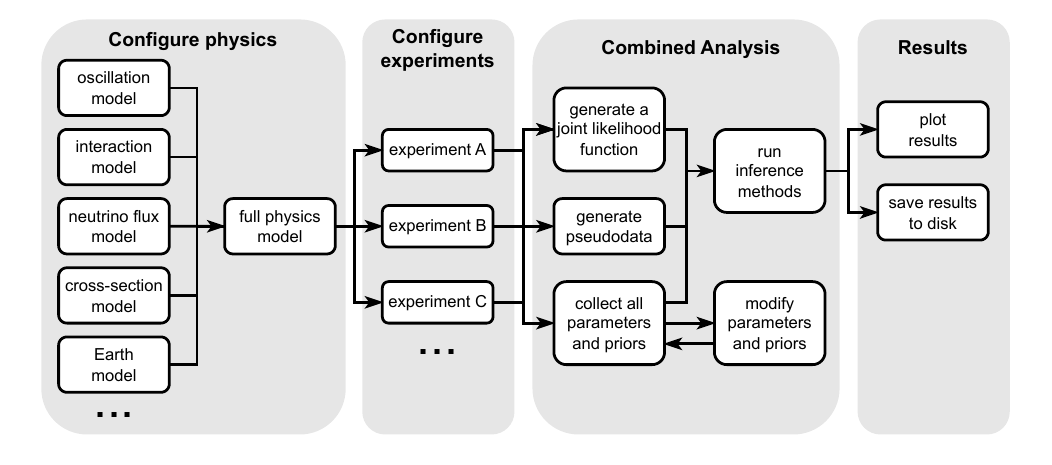}}
\caption{Typical workflow for global analyses of neutrino data with
\emph{Newtrinos.jl}. The layers communicate through two well-defined
interfaces: NamedTuples of parameters and priors, and callable functions
stored in structs. \label{workflow}}
\end{figure}

\section{Availability}\label{availability}

\emph{Newtrinos.jl} is open-source and freely available under the MIT
License at https://github.com/philippeller/Newtrinos.jl .

\section{Related Work}\label{related-work}

Several existing software projects address related but distinct use
cases:

\begin{itemize}
\tightlist
\item
  \textbf{GLoBES} (\citeproc{ref-Huber:2004ka}{Huber et al., 2005},
  \citeproc{ref-Huber:2007ji}{2007}): Simulates long-baseline
  experiments but does not support full global fits.
\item
  \textbf{GAMBIT} (\citeproc{ref-GAMBIT:2017yxo}{Athron \& others,
  2017}): A general-purpose global fitting framework with some support
  for neutrino data, but not tailored for neutrino physics.
\item
  \textbf{PEANUTS} (\citeproc{ref-Gonzalo:2023mdh}{Gonzalo \& Lucente,
  2024}): Focused on solar neutrino modelling.
\item
  \textbf{PhyLiNO} (\citeproc{ref-Hellwig:2025jxe}{Hellwig et al.,
  2025}): A high-performance framework for reactor neutrino data.
\item
  \textbf{PISA} (\citeproc{ref-IceCube:2018ikn}{Aartsen \& others,
  2020}): Designed for atmospheric neutrino analyses.
\end{itemize}

\emph{Newtrinos.jl} complements these efforts by focusing on the
neutrino sector, offering a simple, extensible, and efficient design. It
is currently the only framework in this domain supporting automatic
differentiation. The software has been used in
(\citeproc{ref-Ettengruber:2024fcq}{Ettengruber et al., 2024}),
(\citeproc{ref-Kozynets:2024xgt}{Kozynets et al., 2025}),
(\citeproc{ref-Eller:2025lsh}{Eller et al., 2025}), and
(\citeproc{ref-Eller:2026urd}{Eller, 2026}).

\section{Acknowledgements}\label{acknowledgements}

This work was supported by Germany's Federal Ministry of Research,
Technology and Space (BMFTR) within the ErUM-Data programme under grant
FKZ 05D25PC1 (DEMOS consortium), and partially by the Deutsche
Forschungsgemeinschaft (DFG, German Research Foundation) under Germany's
Excellence Strategy -- EXC-2094/2 -- 390783311, the SFB 1258--
283604770, and NFDI 39/1.

\protect\phantomsection\label{refs}

\end{document}